\documentclass[aps,twocolumn]{revtex4}
\usepackage{graphicx}
\usepackage{ulem}
\begin{document}

\title{Superradiant light scattering from a moving Bose-Einstein condensate}

\author{R. Bonifacio$^{a}$, F. S. Cataliotti$^{b}$, M. Cola$^{a}$, L. Fallani${^c}$,
C. Fort${^c}$, N. Piovella$^{a}$, M. Inguscio$^{c}$}
\affiliation{INFM and European Laboratory for Non-Linear Spectroscopy (LENS)\\
via N. Carrara~1, I-50019 Sesto F.no (FI), Italy\\
a) Dipartimento di Fisica, Universit\`a degli Studi di
Milano, Via Celoria 16, I-20133 Milano, Italy\\
b)Dipartimento di Fisica, Universit\`a di Catania, via S.
Sofia~64, I-95124 Catania, Italy\\
c)Dipartimento di Fisica, Universit\`a di Firenze, via G. Sansone
1, I-50019 Sesto F.no (FI), Italy}

\date{\today}

\begin{abstract}

We investigate the interaction of a moving BEC with a far detuned
laser beam. Superradiant Rayleigh scattering arises from the
spontaneous formation of a matter-wave grating due to the
interference of two wavepackets with different momenta. The system
is described by the CARL-BEC model which is a generalization of
the Gross-Pitaevskii model to include the
\textit{self--consistent} evolution of the scattered field. The
experiment gives evidence of a damping of the matter-wave grating
which depends on the initial velocity of the condensate. We
describe this damping in terms of a phase-diffusion decoherence
process, in good agreement with the experimental results.

\end{abstract}
\maketitle

The experimental realization of Bose-Einstein condensates (BECs)
with alkali trapped atoms has opened the possibility of
investigating several fundamental aspects of quantum mechanics in
macroscopic, i.e. many particle systems \cite{bec:general}. In
superradiant Rayleigh scattering the coherent nature of the
condensate leads to strong correlations between successive
scattering events, as  shown in the pioneering work of Ketterle
and coworkers \cite{inouye99}. This process was then the basis for
the first demonstration of phase-coherent matter wave
amplification \cite{kozuma99}. The effect studied by Ketterle is
an example of a spontaneous formation of a regular density grating
in an atomic system, arising from a collective instability as in
the Collective Atomic Recoil Laser (CARL) \cite{CARL}. In the
absence of thermal broadening (as it happens in a BEC), CARL
appears as a promising source of macroscopically entangled or
number-squeezed atom-atom and/or atom-photon systems
\cite{Moore,Mary,vogels02}. However, in a real BEC several effects
due, for instance, to spontaneous emission, inhomogeneous
broadening and collisions, may seriously inhibit the CARL process
and destroy the coherence in the matter wave field \cite{Burnett}.
The control of decoherence in the photon-BEC interaction would be
a significant step toward the achievement of macroscopic
entanglement of coherent matter waves.

In this paper we investigate both theoretically and experimentally
the influence of the initial velocity of the condensate on
superradiant Rayleigh scattering. In the experiment we produce an
elongated BEC of rubidium atoms and expose it to a single
off-resonant laser pulse directed along the condensate symmetry
axis. The laser is far detuned from any atomic resonance and the
only scattering mechanism present is Rayleigh scattering
\cite{inouye99}. In an elongated condensate a preferential
direction for the scattered photons emerges, causing superradiant
Rayleigh scattering. In this regime the atoms, initially scattered
randomly, interfere with the atoms in the original momentum state
creating a matter-wave grating with the right periodicity to
further scatter the laser photons in the same mode. Both the
matter-wave grating and the scattered light are then coherently
amplified. In our geometry photons are back-scattered with $\vec
k_s\approx -\vec k$, where $\vec k$ is the wave-vector of the
laser photon, and the atoms move away from the original condensate
with a relative momentum $2\hbar k$ in the direction of the laser
beam. The efficiency of the process is limited by the decoherence
between the original and the recoiled atomic wavepackets causing
the damping of the matter-wave grating. We identify two different
mechanisms for decoherence, one resulting from Doppler and mean
field broadening of the matter wave field \cite{inouye99,inouye00}
and the other due to phase diffusion. The latter mechanism,
dependent on the energy separation between the initial and final
states of the system \cite{decoherence,RB}, can be controlled
by initially setting the
condensate into motion. In particular, we observe that
phase diffusion decoherence vanishes when the initial condensate
momentum is such that after the interaction with the laser beam
the scattered atomic wavepacket has the same kinetic energy of the
original condensate in the laboratory frame.

\begin{figure}[t]
\begin{center}
\includegraphics[width=8cm]{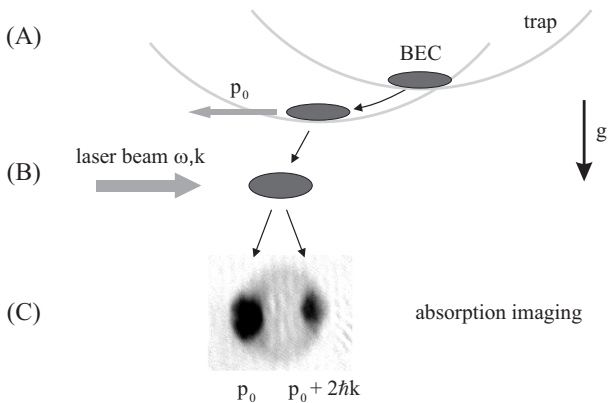}
\caption{Schematics of the experimental procedure. The condensate
is set in motion by a sudden displacement of the magnetic trap
center (A). When the condensate reaches the desired momentum p$_0$
we switch off the magnetic trap and flash the atoms with a far off
resonance laser pulse directed along the condensate symmetry axis
(B). After an expansion time allowing a complete separation of the
momentum components (28 ms) we take an absorption image of the
atoms (C).} \label{schemino}
\end{center}
\end{figure}

The experiment is performed with a cigar-shaped condensate of
$^{87}$Rb produced in a Ioffe-Pritchard magnetic trap by means of
RF-induced evaporative cooling. The axial and radial frequencies
of the trap are $\omega_z / 2 \pi = 8.70(7)$ Hz and $\omega_r / 2
\pi = 90.1(4)$ Hz respectively, with the \textit{z}-axis oriented
horizontally. We tune the atomic velocity after the end of the
evaporative ramp by inducing a collective dipole motion of the
condensate along the \textit{z}-axis. The dipole oscillation is
excited by non-adiabatically displacing the center of the magnetic
trap. When the condensate has reached the maximum velocity in the
magnetic potential, we suddenly switch off the trap and let the
cloud expand with a horizontal velocity proportional to the
displacement of the trap (see Fig.\ref{schemino}). We apply a
square pulse of light along the \textit{z}-axis, 2 ms after the
release of the condensate, when the magnetic field of the trap is
completely switched off, and the atomic cloud has still an
elongated shape. After 2 ms of expansion the radial and axial
sizes of our condensates are typically 10 and 70 $\mu$m,
respectively. The pulse length is controlled with an acousto-optic
modulator. The light comes from a diode laser red-detuned 13 GHz
away from the rubidium D2 line at $\lambda=780$ nm and has an
intensity of 1.35 W/cm$^2$ corresponding to a Rayleigh scattering
rate of roughly 5$\times$10$^2$ s$^{-1}$. The linearly polarized
laser beam is collimated and aligned along the
\textit{z}-axis of
the condensate. In this geometry the superradiant 
light is backscattered and the self-amplified matter-wave
propagates in the same direction of the incident light. In order
to minimize spurious reflections we have aligned the laser beam at
a nonzero angle with respect to the normal to the vacuum cell
windows. After an expansion of 28 ms, when the two momentum
components are spatially separated, we take an absorption image of
the cloud along the horizontal radial direction.

In Fig.~\ref{schemino}C we show a typical absorption image in
which the
left peak is the condensate in its original momentum state $p_0$
and the right peak is formed by atoms recoiling after the
superradiant scattering at $p_0+2\hbar k$. The spherical halo
centered between the two density peaks is due to non-enhanced
spontaneous processes, i.e. random isotropic emission following
the absorption of one laser photon. From a 2D-fit of the pictures
assuming a Thomas-Fermi density distribution we extract the number
of atoms in both the original and the recoiled peaks.

To understand the observed behavior we analyze in detail the
self--consistent interaction between a coherent electromagnetic
wave of amplitude $a$ and a coherent matter wave field $\Psi$.
The evolution of the system is described by the following 1-D
CARL-BEC model, i.e. a
Gross-Pitaevskii model generalized to include the self--consistent
evolution of the scattered radiation amplitude:
\begin{eqnarray}
 i\hbar\frac{\partial\Psi}{\partial t}&=&
 -\frac{\hbar^2}{2m} \frac{\partial^2\Psi}{\partial z^2}
 +\beta N|\Psi|^2\Psi \label{psi} \\
 &+& i\hbar g \left\{a^*e^{i(2kz-\delta t)}- {\rm
c.c.}\right\}\Psi  \nonumber \\
 \frac{da}{dt}&=&gN\int dz|\Psi|^2e^{i(2kz-\delta t)}-\kappa a. \label{a}
\end{eqnarray}
In Eqs.(\ref{psi}) and (\ref{a}), $a(t)=(\epsilon_0
V/2\hbar\omega_s)^{1/2}E_s(t)$ is the dimensionless electric field
amplitude of the scattered beam  with frequency $\omega_s$,
$g=(\Omega/2\Delta)(\omega d^2/2\hbar\epsilon_0 V)^{1/2}$ is the
coupling constant, $\Omega$ is the Rabi frequency of the laser
beam with a frequency $\omega$ detuned from the atomic resonance
frequency $\omega_0$ by $\Delta=\omega-\omega_0$,
$d=\hat\epsilon\cdot\vec d$ is the electric dipole moment of the
atom along the polarization direction $\vec\epsilon$ of the laser,
$V$ is the volume of the condensate, $N$ is the total number of
atoms in the condensate and $\delta=\omega-\omega_s$. The matter
wave field is normalized such that $\int dz |\Psi|^2=1$. The
nonlinear term in Eq.(\ref{psi}) can be neglected since the
experiment has been performed after expansion. The last term in
right-hand side of Eq.(\ref{psi}) represents the self-consistent
optical wave grating, whose amplitude depends on time according to
Eq.(\ref{a}). The first term in the right-hand side of
Eq.(\ref{a}) represents the self--consistent matter-wave grating.
Eq.(\ref{a}) has been written in the ``mean-field'' limit, which
models the propagation effects with a damping term where
$\kappa\approx c/2L$ and $L$ is the condensate length \cite{SR}.

If the condensate is much longer than the radiation wavelength and
approximately homogeneous, then periodic boundary conditions can
be assumed and the wavefunction can be written as
$\Psi(z,t)=\sum_n c_n(t)u_n(z)e^{-in\delta t}$, where
$u_n(z)=(2/\lambda)^{1/2}\exp(2inkz)$ are the momentum eigenstates
with eigenvalues $p_z=n(2\hbar k)$. Introducing the density matrix
$\rho _{m,n}=c_{m}c^{*}_{n}$ and $\omega_{n}=4\omega_R
n^{2}-\delta n$, where $\omega_R=\hbar k^2/2m$ is the recoil
frequency, we obtain from Eqs.(\ref{psi}) and (\ref{a}):
\begin{widetext}
\begin{eqnarray}
\frac{d \rho_{m,n}}{d t}&=& -i(\omega_{m}-\omega_{n})\rho_{m,n}+g\{
a(\rho_{m,n-1}-\rho_{m+1,n})+a^{*}(\rho_{m-1,n}-\rho_{m,n+1})\}
 -\frac{\tau}{2}(\omega_{m}-\omega_{n})^{2}\rho_{m,n} \label{mn}\\
\frac{da}{dt} &=& g N \sum_{n}\rho_{n,n+1}-\kappa a\label {a:2}.
\end{eqnarray}
\end{widetext}
The last term added in Eq.(\ref{mn}) describes a phase-diffusion
decoherence process, whose amplitude is characterized by a
constant $\tau$. This term, fundamental to describe our
experimental results, arises from a $\delta$-correlated gaussian
noise on the eigenenergies of the system and
causes
the decay of the off-diagonal matrix elements, so that the density
matrix becomes diagonal in the basis of the recoil momentum
states. Eq.(\ref{mn}) may be written as a master equation
\cite{Milburn} for the density operator $\hat\rho=\sum_{m,n}\rho
_{m,n}|m\rangle \langle n|$:
\begin{equation}
\frac{d\hat\rho}{dt}=-\frac{i}{\hbar}[\hat H,\hat\rho]-\frac{\tau}{2}
[\hat H_{0},[\hat H_{0},\hat\rho]],
\label{ME}
\end{equation}
where $\hat H=\hat H_{0}+\hat V$, $\hat
H_{0}=4\hbar\omega_{R}\hat{p}^{2}-\hbar\delta \hat{p}$, $\hat
V=i\hbar g(a^{*}e^{2ikz}-h.c.)$ and $\hat p=p_z/2\hbar k$ is the
normalized momentum operator with eigenstates $|n\rangle$ and
eigenvalues $n$. The phase destroying term with the double
commutator of the Lindblad form in the right-hand side of
Eq.(\ref{ME}) generates the damping term added in Eq.(\ref{mn}).
It has appeared in many models of decoherence and induces
diffusion in variables that do not commute with the Hamiltonian,
preserving the number of atoms in the condensate. In this term we
have neglected the interaction $\hat V$ in the weak-coupling limit
$g^2N/\kappa\ll\omega_R$.

\begin{figure}[t]
\begin{center}
\includegraphics[width=8cm]{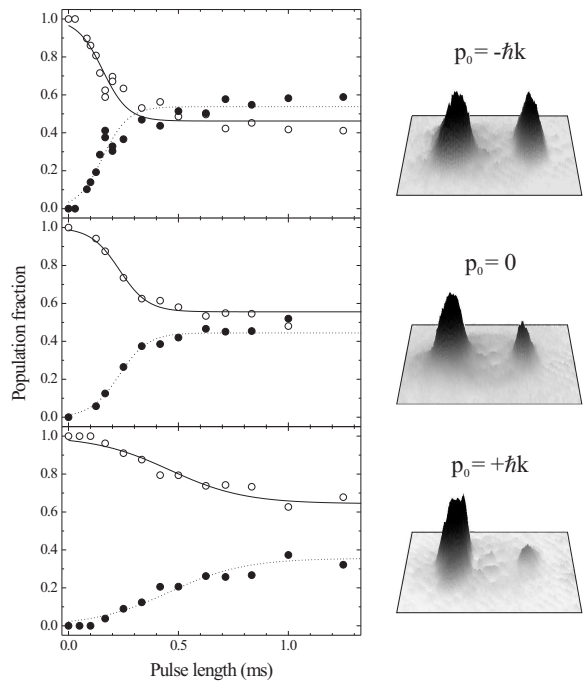}
\caption{ Left) Time evolution of population in the original
condensate (empty circles) and in the recoiled wavepacket (filled
circles) for different pulse durations. The solid line is a fit
with the hyperbolic tangent (\ref{tanh}) predicted by the
theoretical model, the dotted line is just one minus the fit
curve. The momentum of the original condensate is set to $-\hbar
k$ (top), $0$ (center) and $+\hbar k$ (bottom). Right) Plot of the
atomic density profile after interaction with a $250 \mu s$ pulse
for the three cases of original momentum as on the left. The laser
detuning and intensity are 13 GHz and 1.35 W/cm$^2$ respectively.}
\label{time}
\end{center}
\end{figure}

In our experimental conditions the superradiant Rayleigh
scattering involves only neighboring momentum states, i.e.
transitions from the initial momentum state $p_{0}=n(2\hbar k)$ to
the final momentum state $(n+1)2\hbar k$. In this limit, our
system is equivalent to a two-level system and Eqs.(\ref{mn}) and
(\ref{a:2}) reduce to a set of equations formally equivalent to
the well-known Maxwell-Bloch system \cite{inouye00,CARLQ}:
\begin{eqnarray}
\frac{dS}{dt}&=&gAW \label{MB1}-\gamma_n S\\
\frac{dW}{dt}&=&-2g(AS^{*} +h.c.)  \label{MB2}\\
\frac{dA}{dt}&=&gN S-(\kappa -i\Delta_n)A \label{MB3},
\end{eqnarray}
where $S=\rho_{n,n+1}e^{i\Delta_n t}$, $A=a e^{i\Delta_n t}$,
$W=P_n-P_{n+1}$ is the population fraction difference between the
two states (where $P_n=\rho_{n,n}$ and $P_n+P_{n+1}=1$),
$\Delta_{n}=\omega-\omega_{s}-4\omega_{R}(2n+1)$ and the
decoherence rate $\gamma_n$ is given by:
\begin{equation}
\gamma_n = \gamma_0 + \frac{\tau}{2} \Delta_{n}^{2}= \gamma_0 +
\frac{\tau}{2}\left[\omega-\omega_{s}-4\omega_{R}\left(\frac{p_0}{\hbar k}+1\right)\right]^{2}.
\label{dec}
\end{equation}
To the decoherence rate $\gamma_n$ we have added an extra term
$\gamma_0$ taking into account other coherence decay mechanisms,
as for instance Doppler and inhomogeneous broadenings of the
two-photon Bragg resonance \cite{inouye99,inouye00}. We note that
in Eq.(\ref {MB3}), $S$ represents half of the amplitude of the
matter-wave grating. In fact, if $\Psi\approx
c_nu_n(z)+c_{n+1}u_{n+1}(z)$, the longitudinal density is
$|\Psi|^2\approx(2/\lambda)\{1+2{\rm Re}[S^*\exp(2ikz-i\Delta_n t)]\}$, which
describes a matter wave grating with a periodicity of half the
laser wavelength. The main result is that the second term of
Eq.(\ref{dec}), arising from a phase diffusion decoherence
mechanism, depends on the frequency detuning between the incident
and scattered radiation beams and on the initial momentum of the
condensate, $p_0=n(2\hbar k)$. We observe that the
velocity-dependent term of the decoherence rate is invariant under
Galilean transformation. In fact, in a frame moving with respect
to the laboratory frame, the shift of $p_{0}$ compensates the
Doppler shift of the frequency difference $\omega-\omega_s$.

\begin{figure}[t]
\begin{center}
\includegraphics[width=7cm]{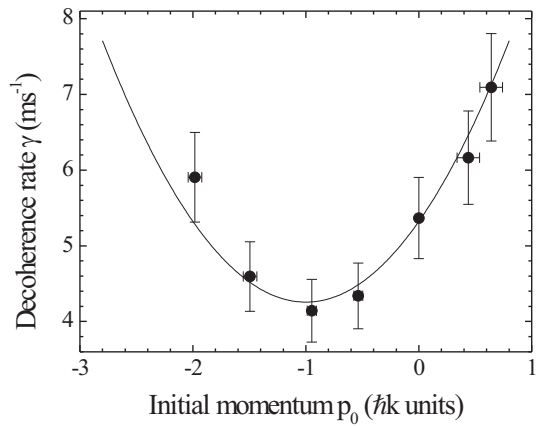}
\end{center}
\caption{Decoherence rate as a function of the initial momentum of
the condensate. The solid line is a fit of the experimental data
with a parabola centered in p$_0=-\hbar k$, as expected from the
theoretical model.} \label{parabola}
\end{figure}

Our experimental conditions match those for the superfluorescent
regime \cite{SF}, in which the field loss rate $\kappa$ is much
larger than the coupling rate $g\sqrt N$.
In this regime, for $t\gg\kappa^{-1}$, we can perform the adiabatic
elimination $A \simeq gNS/(\kappa-i\Delta_n)$. The analytical
solution for the fraction of atoms with initial momentum
$p_0=n(2\hbar k)$ is
\begin{eqnarray}
P_n &=&1-\frac{1}{2}\left(1-\frac{2\gamma_n}{G}\right)\times\nonumber\\
&\times&\left\{1+\tanh\left[\left(G - 2\gamma_n\right)(t-t_0)/2\right]\right\},
\label{tanh}
\end{eqnarray}
where $G=2g^2N\kappa/(\kappa^{2}+\Delta_{n}^2)$ is the
superradiant gain and $t_0$ is a delay time. In our experiment
$\kappa\gg\Delta_n$, so that $G\approx 2g^2N/\kappa$ hence
independent from the atomic velocity.
Eq.(\ref{tanh}) assumes the threshold condition $G>2\gamma_n$ i.e.
the gain must be larger than the decoherence rate. Fig.\ref{time}
reports the population fractions of the initial wavepacket, $P_n$,
and of the scattered wavepacket, $P_{n+1}$, as functions of the
laser pulse duration for three different initial momenta $p_0$.
>From the fits we extract the values of $G$ and $\gamma_n$ for
different $p_0$. The measured value of $G=19(3)$ ms$^{-1}$ does
not appreciably depend on $p_0$. On the contrary we observe a
strong dependence of the decoherence rate $\gamma_n$ on the
initial momentum $p_0$.
In Fig.\ref{parabola} we plot the experimental points for the
decoherence rate $\gamma_n$ as a function of the initial momentum
of the atoms. The data show a parabolic behavior in good agreement
with the prediction of Eq.(\ref{dec}) if one assumes
$\omega=\omega_{s}$ in the laboratory frame, with fit values
$\gamma_0$=4.2(2) ms$^{-1}$ and $\tau$=2.4(2)$\cdot10^{-7}$s. The
theoretical calculation of the linewidth of the Bragg resonance
\cite{stenger} for our experimental parameters predicts a value
$\gamma_0\approx$ 3 ms$^{-1}$ close to the value obtained from the
fit in Fig.\ref{parabola}. Notice that the decoherence rate is
minimized for $p_0=-\hbar k$. Indeed, if the initial momentum is
$-\hbar k$, after scattering the atoms have the same kinetic
energy and, with the above assumption for the scattered light
frequency $\omega_s$, the phase-diffusion decoherence term in
Eq.(\ref{dec}) is zero. This identifies a subspace which is
decoherence free with respect to the phase destroying process
\cite{df}.

In our experimental apparatus it is very difficult to completely
exclude the presence of pump light diffused by the vacuum cell
windows. We have measured the diffused light present in our
chamber in the direction opposite to the pump beam to be $1.0(3)
\times 10^{-5}$ of the pump beam intensity. This amount of light
is of the same order of magnitude of the equivalent input noise of
the superradiant process \cite{notanoise}. This justifies the
assumption that $\omega_s=\omega$ used to fit the experimental
data of Fig. \ref{parabola}. We remark that this back diffused
light can not explain our results in terms of Bragg scattering since
the width of the Bragg resonance is one order of magnitude smaller
than the range of momenta explored in our experiment and shown in
Fig. \ref{parabola}. Furthermore the hyperbolic tangent dependence
of the atomic population in Fig. \ref{time} can only be explained
by the self consistent amplification of the matter wave grating
and of the backscattered light as described in Eqs.(\ref{psi}) and (\ref{a}).

In conclusion, we have studied the superradiant light scattering
from a moving Bose-Einstein condensate. The efficiency of the
overall process is fundamentally limited by the decoherence
between the two atomic momentum states.
With the assumption for the scattered light frequency
$\omega_s=\omega$ in the laboratory frame, Eq.(\ref{dec}) predicts
the parabolic behavior of the decoherence rate as a function of
the initial momentum, in agreement with the experimental results
as shown in Fig.\ref{parabola}. The fully quantized version of the
CARL-BEC model offers the possibility of investigating the
realisation of macroscopic atom-atom or atom-photon entanglement
\cite{Moore, Mary}. The control of decoherence obtained in this
work represents a significant step in this direction.

This work has been supported by the EU, INFM and MIUR. We thank B.
Englert and S. Olivares for useful discussions.

\end{document}